\documentclass[3p,times]{elsarticle}

\usepackage{ecrc}

\volume{00}

\firstpage{1}

\journalname{Annals of Physics}

\runauth{Saha, Gefen, Burmistrov, Shnirman and Altland}

\jnltitlelogo{Annals of Physics}

\usepackage{epsfig,amssymb,amsmath,latexsym}
\usepackage{subfigure}
\usepackage{wrapfig}
\usepackage{float}
\usepackage{color}
\newcommand\beq{\begin{equation}}
\newcommand\eeq{\end{equation}}

\newcommand\bea{\begin{eqnarray}}
\newcommand\eea{\end{eqnarray}}

\newcommand\bi{\begin{itemize}}
\newcommand\ei{\end{itemize}}

\newcommand\ie{{\it{i.e.}}}

\newcommand\eg{{\it{e.g.}}}

\newcommand\qd{{\textsf{QD}}}
\newcommand\qdd{{\textsf{QD~}}}
\newcommand\qds{{\textsf{QDs}}}
\newcommand\qdsd{{\textsf{QDs~}}}

\newcommand\ee{{\textsf{e-e}}}
\newcommand\eed{{\textsf{e-e~}}}

\newcommand\hs{{\textsf{HS}}}
\newcommand\hsd{{\textsf{HS~}}}

\newcommand\mf{{\textsf{B}}}
\newcommand\mfd{{\textsf{B~}}}

\newif\ifboo \boofalse

\usepackage[figuresright]{rotating}

\begin{document}

\begin{frontmatter}



\dochead{}

\title{A quantum dot close to Stoner instability: the role of Berry's Phase}


\author{ARIJIT SAHA AND YUVAL GEFEN}

\address{Department of Condensed Matter Physics, Weizmann Institute of Science, Rehovot 76100, Israel}

\author{IGOR BURMISTROV}

\address{Landau Institute for Theoretical Physics, 119334 Moscow, Russia}

\author{ALEXANDER SHNIRMAN}

\address{Institut f\"{u}r Theorie der Kondensierten Materie, Karlsruhe Institute of Technology, 76128 Karlsruhe, 
Germany\\
\& DFG Center for Functional Nanostructures (CFN), Karlsruhe Institute of Technology, 76128 Karlsruhe, Germany}

\author{ALEXANDER ALTLAND}

\address{Institut f\"{u}r Theoretische Physik, Universit\"{a}t zu K\"{o}ln, D-50973 K\"{o}ln, Germany}

\begin{abstract}
The physics of a quantum dot with electron-electron interactions is well captured by 
the so called "Universal Hamiltonian" if the dimensionless conductance of the dot is much higher than unity.
Within this scheme interactions are represented by three spatially independent terms which describe 
the charging energy, the spin-exchange and the interaction in the Cooper channel. In this paper we concentrate 
on the exchange interaction and generalize the functional bosonization formalism developed earlier for the 
charging energy. This turned out to be challenging as the effective bosonic action is formulated in terms of a 
vector field and is non-abelian due to the non-commutativity of the spin operators. Here we develop a 
geometric approach which is particularly useful in the mesoscopic Stoner regime, i.e., when the strong 
exchange interaction renders the system close the the Stoner instability. We show that it is sufficient to sum over 
the adiabatic paths of the bosonic vector field and, for these paths, the crucial role is played by the Berry phase.
Using these results we were able to calculate the magnetic susceptibility of the dot. The latter, in close vicinity 
of the Stoner instability point, matches very well with the exact solution (Pis'ma v ZhETF {\bf{92}}, 202 (2010)). 
\end{abstract}


\end{frontmatter}


\section{Introduction}
\label{secI}
Over the past few decades physics of quantum dots (\qds) has become a focal point of 
research in nanoelectronics. The introduction of the "Universal Hamiltonian`` 
~\cite{PhysRevB.62.14886,AleinerBrouwerGlazman,alhassidreview,KamenevAndreev2} has 
made it possible to take into account the effects of 
electron-electron (\ee) interaction within a quantum dot (\qd) in a controlled way. 
This approach is applicable for a normal-metal \qd\ in the 
metallic regime when the Thouless energy $E_{Th}$ and the mean single particle level spacing 
$\delta$ satisfy $g \equiv E_{Th}/\delta \gg 1$ ($g$ is the dimensionless conductance).  Within this scheme 
interactions are split into a sum of three spatially independent contributions in the charging, 
spin-exchange, and Cooper channels. The charging term leads to the phenomenon of Coulomb blockade, 
while the spin-exchange term can drive the system towards the Stoner instability~\cite{Stoner}.

In bulk systems the exchange interaction competes with the kinetic energy leading to Stoner instability. 
In finite size systems mesoscopic Stoner regime may be a precursor of bulk thermodynamic 
Stoner Instability~\cite{PhysRevB.62.14886}. More precisely,
one distinguishes three regimes depending on the strength of the exchange interaction:
(a)~a phase with the total spin of the dot equal zero, (b)~the mesoscopic Stoner regime in which the total spin of the dot is finite but not proportional to the volume of the dot, and (c)~the thermodynamic ferromagnetic phase where magnetization is proportional to the volume. The mesoscopic Stoner regime can be realized in \qdsd 
made of materials close to the thermodynamic Stoner instability, \eg, Co impurities in Pd or Pt host, 
Fe dissolved in various transition metal alloys, Ni impurities in Pd host, and Co in Fe grains, as well as new 
nearly ferromagnetic rare earth materials~\cite{PGambardellaetal,Mpourmpakisetal,PCCanfield}. 

Notably, the inclusion of the spin-exchange turned out to be non-trivial as the resulting path integral 
action is non-Abelian~\cite{KiselevGefen,PismaJETP.92.202}.
To understand the complexity of the problem we compare with the case when only the charging interaction is taken 
into account~\cite{KamenevGefen,EfetovTschersich,Lerner}. It was suggested by Kamenev and Gefen~\cite{KamenevGefen} to 
take the following steps in solving that problem: (a)~start from a fermionic action which includes an \eed interaction 
term quartic in the fermionic Grassman variables, (b)~perform a Hubbard-Stratonovich (\hs) transformation by introducing 
an auxiliary bosonic field, (c)~perform a gauge transformation over the Grassman variables which makes all the non-zero 
Matsubara components of the (\hs) field decouple from the fermionic fields in the action, and, finally, (d)~integrate out 
the fermions. The resulting, purely bosonic action, is quadratic in the bosonic non-zero Matsubara components, which 
renders the problem easily solvable. The trick of gauge-integrating over Grassman variables does not 
work for the non-abelian case~\cite{KiselevGefen}, so that an alternative approach is needed.
\begin{figure}
\begin{center}
\includegraphics[width=12.0cm,height=7.0cm]{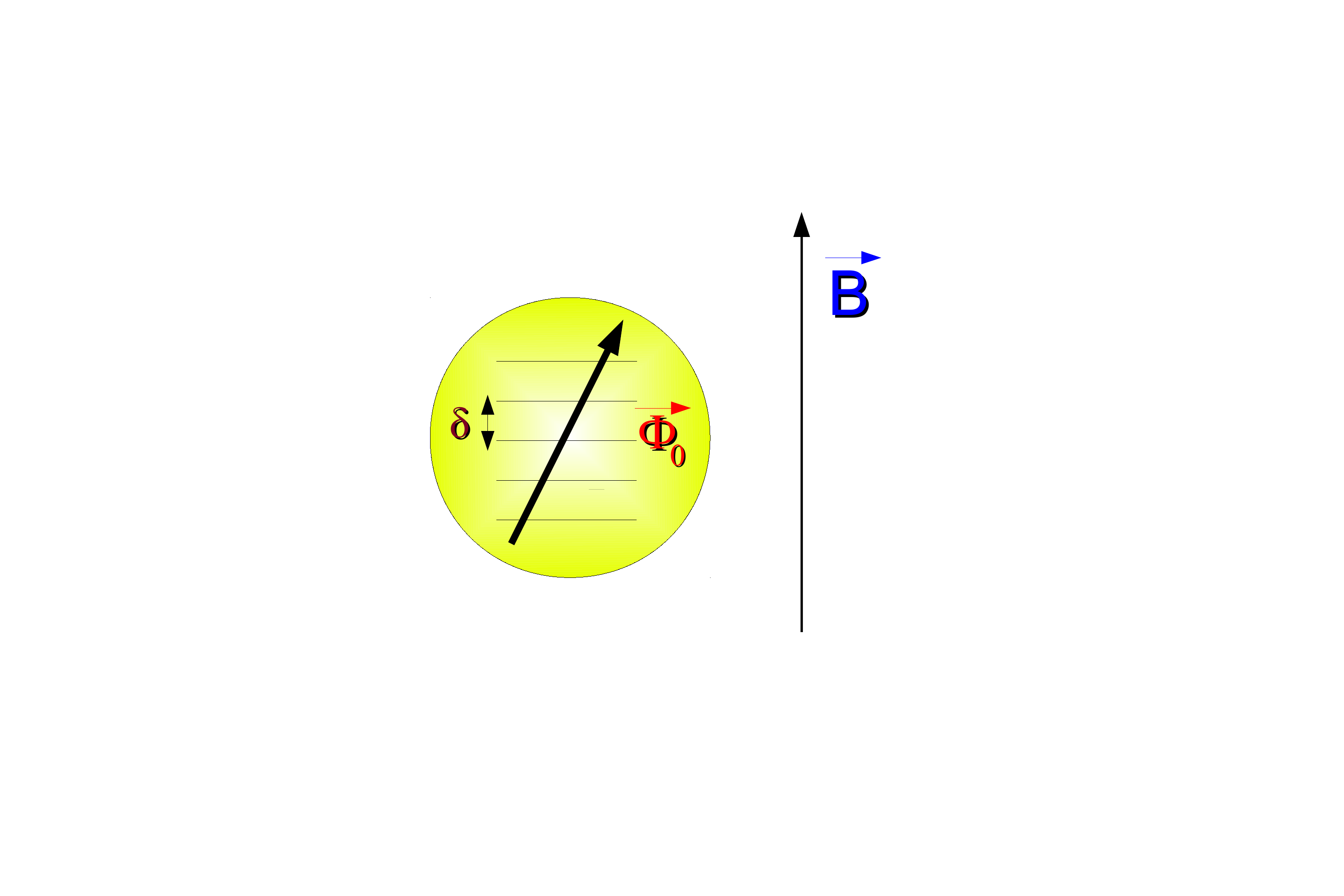}
\caption{(Color online) Cartoon scheme of an isolated \qdd. Here $\delta$ is the single particle level spacing,
$\Phi_{0}$ is the zero component of the auxiliary \hsd vector bosonic field, and \mfd is the applied 
magnetic field.}
\label{figgeometry}
\end{center}
\end{figure}

There have been several attempts to account for charge and spin interactions in a \qdd including a 
rate equation analysis~\cite{AlhassidRupp,baranger} and a perturbative
expansion~\cite{KiselevGefen}. Alhassid and Rupp~\cite{AlhassidRupp} have analyzed some 
aspects of the problem exactly. More recently an exact solution of the isotropic spin interaction model based on 
the generalized Wei-Norman-Kolokolov method~\cite{IVKolokolov} has been presented~\cite{PismaJETP.92.202}.
In this exact solution several observables, including the tunneling density of states and magnetic 
susceptibility have been calculated below the Stoner instability point for an equidistant spectrum.
The effects of disorder have been addressed in Ref~\cite{BurmistrovDisorder}. 
The tunneling density of states exhibits a non-monotonous behavior as a function of energy, and the 
magnetic susceptibility emerges out to be a sum of Pauli and Curie like terms. 

In this article we present an approximative geometric approach to tackle the isotropic spin-exchange model.
Our results are in agreement with the exact results ~\cite{PismaJETP.92.202}
for the partition function and the magnetic susceptibility within the mesoscopic Stoner instability regime. 
Our rationale behind developing an approximation scheme, given the exact solution, 
is the high complexity and inflexibility of the exact method. We thus expect to be able to apply 
our geometric approach in cases where the exact method is inapplicable or too complicated. These should cover a broad 
range of problems involving spin transport through \qdd coupled to normal leads and through an array of \qdsd.

This paper is organized as follows. In Sec.\ref{secII} we consider an isolated \qdd with isotropic exchange interaction, 
and derive an effective action in terms of an auxiliary \hsd bosonic vector field. In Sec.\ref{secIII} we perform a 
perturbative expansion of our effective action in powers of the Berry's connection operator, and show that in the 
lowest order of this expansion the Berry phase governs the effective action. In Sec.\ref{secIV} we calculate the partition 
function and the magnetic susceptibility of the \qdd, and discuss the effect of Berry phase on susceptibility. 
Sec.\ref{secV} contains a summary of our analysis. In the appendices (A, B and C) we present alternative methods 
of calculation, which provide further justification of our results.

\section{Hamiltonian and Effective action}
\label{secII}
\noindent
{\it{In this section we perform a \hsd transformation and obtain the effective action in terms of an auxiliary \hsd bosonic 
vector field $\vec \Phi$. We, then perform a unitary rotation $R$ to the instantaneous direction of $\vec \Phi$ and 
rewrite the effective action in terms of the Berry connection operator $R^{-1}\dot{R}$.
Finally, integrating out the fermions we obtain the effective action of the isolated \qdd in terms of the zero
component of the \hsd field and $R^{-1}\dot{R}$.}}\\

A quantum dot in the metallic regime, $g\gg 1$, is described by the universal Hamiltonian~\cite{PhysRevB.62.14886}:
\begin{equation}
H =H_0 + H_C+H_J+H_{\lambda} . \label{EqUnivHam_1.1}
\end{equation}
The noninteracting part of the universal Hamiltonian reads
\begin{equation}
H_0= \sum\limits_{\alpha,\sigma} \epsilon_{\alpha} a^\dag_{\alpha,\sigma} a^{\phantom \dag}_{\alpha,\sigma}\ ,\label{EqUnivHam_1.2}
\end{equation}
where $\epsilon_{\alpha}$ denotes the energy of a spin-degenerate (index $\sigma$) single 
particle level $\alpha$. The charging interaction term
\begin{equation}
H_C=E_C \left ( \hat{N} - N_0\right )^2
\end{equation}
accounts for the Coulomb blockade. Here, $E_C$ denotes the charging energy of the \qd, $N_0$ represents the background charge, and
$\hat N =\sum_{\alpha,\sigma} a^\dag_{\alpha,\sigma} a^{\phantom \dag}_{\alpha,\sigma}$ is the operator of the total number of electrons of the dot. For the isolated \qd~  the total number of electrons is fixed and, therefore, the charging interaction term can be omitted. The term
\begin{equation}
H_J = -J {\bf \hat S}^2
\end{equation}
represents the ferromagnetic ($J>0$) exchange interaction within the dot where ${\bf \hat S}=
\sum_{\alpha} a^\dag_{\alpha,\sigma_{1}}{\bf S}_{\sigma_{1}\sigma_{2}}a^{\phantom \dag}_{\alpha,\sigma_{2}}$ is 
the operator of the total spin of the dot. Here ${\bf S}_{\sigma_{1}\sigma_{2}}
\equiv(1/2) \vec \sigma_{\sigma_{1}\sigma_{2}}$, where $\vec{\sigma}=(\sigma_{x}, \sigma_{y}, 
\sigma_{z})$ is a vector made of Pauli matrices. The interaction in the Cooper channel is described by 
\begin{equation}
H_\lambda = \lambda T^{\dag} T^{\phantom \dag}, \qquad T = \sum_{\alpha} a^{\phantom\dag}_{\alpha,\uparrow} a^{\phantom \dag}_{\alpha,\downarrow} .
\end{equation}
In what follows we do not take into account $H_\lambda$ for the following reasons. For the dots fabricated 
in 2D electron gas the interaction in the Cooper channel is typically repulsive and, therefore, renormalizes 
to zero~\cite{AleinerBrouwerGlazman}. In the case of 3D quantum dots realized as small metallic grains, the interaction 
in the Cooper channel can be attractive, giving rise to interesting competition between superconductivity and 
ferromagnetism~\cite{Schechter,AlhassidSF1,AlhassidSF2}. In that case we assume that there is a weak magnetic field 
which suppresses the Cooper channel. 

As explained above we restrict ourselves to a simplified version of the universal Hamiltonian, where the interaction in 
the charging and Cooper channel is set to zero (more precisely, the charging energy is fixed because the number of particles 
is fixed):
\begin{eqnarray}
\label{eq:HJ}
H = \sum_{\alpha,\sigma} \epsilon_\alpha  a^\dag_{\alpha,\sigma} a^{\phantom \dag}_{\alpha,\sigma} - 
J {\bf \hat S}^2\ .
\end{eqnarray}

Our aim is to calculate the partition function ${\mathcal {Z}} = \int D\bar \Psi D\Psi \,\exp{[{\cal S}_\Psi]}$, 
where the imaginary time action is given by 
\begin{eqnarray}
&&{\cal S}_\Psi = \int\limits_0^\beta {\cal L}d\tau\
= \int\limits_0^\beta d\tau\, \Big[\sum_{\alpha}{\bar\Psi}_{\alpha}
(-\partial_\tau  + \mu) \Psi_{\alpha} - H\Big] \nonumber \\
&&= \int\limits_0^\beta d\tau \left[\sum_{\alpha\sigma} \bar\psi_{\alpha\sigma}
(-\partial_\tau -\epsilon_\alpha +\mu) \psi_{\alpha\sigma} + J  
\left[\sum_{\alpha\sigma_1 \sigma_2} \bar \psi_{\alpha\sigma_1} 
{\bf S}_{\sigma_1 \sigma_2} \psi_{\alpha\sigma_2} \right]^2\right]\ .
\label{Inv1}
\end{eqnarray}
Here $\mu$ is the chemical potential, $\beta \equiv 1/T$, $T$ the temperature, and we have introduced the Grassmann variables
${\bar\Psi}_{\alpha} = ({\bar\psi}_{\alpha \uparrow}, {\bar\psi}_{\alpha \downarrow})^{T}$, 
$\Psi_{\alpha} = (\psi_{\alpha \uparrow}, \psi_{\alpha \downarrow})$ to represent electrons on the \qd.

In Eq.\ref{Inv1}, the exchange energy is quartic in the fermionic fields. Hence, we can perform a \hsd transformation to 
obtain an effective action quadratic in the fermionic fields, with an auxiliary vector bosonic field 
$\vec{\Phi}(\tau)=(\Phi_{x}(\tau), \Phi_{y}(\tau), \Phi_{z}(\tau))$. The effective action reads 
\begin{eqnarray}
&&{\cal S}_{\Psi,\Phi} = \int\limits_0^\beta d\tau \left[\sum_{\alpha} \bar\Psi_{\alpha}
\left(-\partial_\tau -\epsilon_\alpha +\mu - \vec \Phi\cdot \vec{\bf S}\right)\Psi_{\alpha} 
-\frac{|\vec \Phi |^2}{4J} \right]\ .
\label{Inv2}
\end{eqnarray}
Integrating out the fermions, we obtain the effective action in terms of 
the auxiliary vector field $\vec{\Phi}$ only 
\begin{eqnarray}
&& {\cal S}_{\Phi} = \sum_{\alpha}\mathrm{tr\;ln}\left(-\partial_\tau - \epsilon_{\alpha} + \mu - 
\vec{\Phi}\cdot\vec{\bf S}\right) - \int\limits_0^\beta d\tau \frac{|\vec \Phi |^2}{4J} \nonumber \\
&&= \sum_{\alpha}\mathrm{tr\;ln}\left(-\partial_\tau - \epsilon_{\alpha} + \mu -
\Phi (\tau){\vec n}(\tau)\cdot\frac{\vec{\sigma}}{2}\right) - \int\limits_0^\beta d\tau\, \frac{\Phi^2(\tau)}{4J} \ .
\label{Inv3}
\end{eqnarray}
Here ${\vec n}(\tau)$ is a unit vector along the direction of
$\vec \Phi(\tau)$ and $\Phi(\tau) \equiv |\vec \Phi(\tau)|$. The first part of the action in Eq.\ref{Inv3} describes the coupling of non-interacting electrons to a time varying magnetic field (exchange field) of magnitude $\Phi(\tau)$ and direction ${\vec n}(\tau)$.
The resulting bosonic action in Eq.\ref{Inv3} is non-abelian due to the non commutativity of the 
Pauli matrices. Note that the problem is 'isotropic' in the sense that the outcome should not depend on 
the initial and final direction of ${\vec n}$.

We are guided by the idea that close to Stoner instability the amplitude of the exchange field
$\Phi$ is large, \ie, a large total spin develops on the dot ($\vec \Phi \sim J \bf S$).  In 
that situation one can distinguish between the adiabatic and the non-adiabatic paths of $\vec \Phi(\tau)$. 
The former involve Matsubara frequencies such that $|\omega_m| \ll \Phi $. We argue that the non-adiabatic paths 
do not contribute considerably to the partition function (except for providing for proper normalization), since 
the electrons do not manage to react to the fast changes of $\vec\Phi$. 
Hence, we concentrate on the adiabatic paths and perform an expansion in the time variation of 
${\vec n}$. We transform to a coordinate system in which $\vec n$ coincides with the $z$-axis 
\begin{eqnarray}
&& {\vec n}\cdot\vec{\sigma} =  R \sigma_z R^{-1},
\label{Inv4}
\end{eqnarray}
where $R$ is a unitary rotation matrix. Eq.~(\ref{Inv4}) identifies $R$ as an element of ${\rm SU(2)/U(1)}$.
Indeed, if we employ the Euler angle representation
\begin{eqnarray}
&& R = \exp{\left[-{i\phi\over 2} \sigma_z\right]} \exp{\left[-{i\theta\over 2} \sigma_y\right]} \exp{\left[-{i\psi\over 2} \sigma_z\right]}\ ,
\label{Inv5}
\end{eqnarray}
then the angles $\phi$ and $\theta$ determine the direction of $\vec n$, while $\psi$ is arbitrary, i.e., 
the condition (\ref{Inv4}) is achieved with any value of $\psi$. Thus, $\psi$ represents the 
gauge freedom of the problem. We obtain 
\begin{eqnarray}
\sum_{\alpha}\mathrm{tr\;ln}\left(-\partial_\tau - \epsilon_{\alpha} + \mu -
\Phi (\tau){\vec n}(\tau)\cdot\frac{\vec{\sigma}}{2}\right)=\sum_{\alpha}\mathrm{tr\;ln}\left(-\partial_\tau - \epsilon_{\alpha} + \mu - \Phi(\tau)\,\frac{\sigma_z}{2} -
R^{-1}\partial_\tau R\right)\ .
\label{Inv6}
\end{eqnarray}
In the transformation (\ref{Inv6}) one can, in the spirit of Ref.~\cite{KamenevGefen}, think of $R$ as being applied to the 
fermionic field. That is, one first introduces $\Psi'_\alpha$ via $\Psi_\alpha = R \Psi'_\alpha$ 
and, then, integrates over $\Psi'_\alpha$. In this case it is convenient to choose the 
gauge $\psi$ so that $R$ remains periodic in Matsubara time upon a continuous change of $\phi$ to $\phi+2\pi$. 
Then $\Psi'_\alpha$ is anti-periodic as it should be. This can be achieved, e.g., by fixing the gauge as 
$\psi(\tau) = - \phi(\tau)$. In what follows we work in this gauge. 

Next, we represent the amplitude of the exchange field as a sum of its zero frequency component 
and of the rest, $\Phi(\tau)=\Phi_0 + \delta\Phi(\tau)$, such that $\int\limits_0^\beta \delta\Phi(\tau) d\tau=0$. 
It is clear and can be easily checked that adiabatic longitudinal fluctuations $\delta \Phi$ do not contribute substantially 
to the effective action (see~\ref{appendixAA} for a more formal treatment). Thus, we disregard that part of 
$\delta \vec{\Phi}$ and obtain
\begin{eqnarray}
\sum_{\alpha}\mathrm{tr\;ln}\left(-\partial_\tau - \epsilon_{\alpha} + \mu - \Phi(\tau)\,\frac{\sigma_z}{2} -
R^{-1}\partial_\tau R\right) \approx \sum_{\alpha}\mathrm{tr\;ln}\left(-\partial_\tau - \epsilon_{\alpha} + \mu - \Phi_0\,\frac{\sigma_z}{2} -
\eta(\tau) \right)\ ,
\label{Inv6a}
\end{eqnarray}
where 
\begin{eqnarray}
\eta(\tau)&\equiv& R^{-1}\left(\partial_\tau R \right)  \nonumber\\
&=&-{i\over 2} \dot \phi (\cos \theta-1) \sigma_z +{i\over 2}\,\dot \phi \left[\sin\theta (\cos\phi\sigma_x+\sin\phi\sigma_y)\right] 
-{i\over 2}\, \dot \theta \left[\cos\phi \sigma_y - \sin\phi \sigma_x\right] \nonumber\\
\label{Inv7}
\end{eqnarray}
is the Berry's connection operator in the gauge $\psi(\tau) = -\phi(\tau)$.

\section{Expansion in powers of the Berry connection $R^{-1}\dot{R}$}
\label{secIII}
\noindent
{\it{In this section we perform a perturbation expansion in powers of the operator $\eta$,
and obtain an effective action which encompasses both the longitudinal 
fluctuations and the Berry phase term.}}\\

We aim at the expansion of the action (\ref{Inv3}) in powers
of the Berry connection operator $\eta = R^{-1}\dot{R}$ (see~\ref{Inv7}). 
It is expected that this expansion quickly converges for the adiabatic 
paths of $\vec \Phi(\tau)$. We write ${\cal S}_{\Phi} = \sum_{n} {\cal S}_{\Phi}^{(n)}$.  
The zeroth order term ${\cal S}_{\Phi} ^{(0)}$ can be obtained by calculating the grand canonical potential 
of the noninteracting electrons  subject to a constant Zeeman field $\Phi_0$. 
We obtain

\begin{eqnarray}
{\cal S}_{\Phi} ^{(0)}= \sum_{\alpha}\mathrm{tr\;ln}\left(-\partial_\tau - \epsilon_{\alpha} + \mu -
\Phi_0 \,\frac{\sigma_z}{2}\right) - \int\limits_0^\beta d\tau\, \frac{\Phi^2(\tau)}{4J}
=  -\beta \Omega_0(\Phi_0)  - \frac{\beta \Phi_0^2}{4J} - \sum_{m\neq 0} \frac{\beta}{4J}\, \delta\Phi_m \delta\Phi_{-m}\ ,
\label{eq:S0}
\end{eqnarray}
where
\begin{equation}
-\beta \Omega_0(\Phi_0) = \ln Z_0 = \sum_\alpha \left[ \ln \left(1+ e^{-\beta\left(\epsilon_\alpha - \frac{\Phi_0}{2}-\mu\right)}\right)
+\ln \left(1+ e^{-\beta\left(\epsilon_\alpha + \frac{\Phi_0}{2}-\mu\right)}\right)
\right]\ .
\end{equation}
Here $Z_0$ is the partition function of noninteracting electrons subject to a magnetic 
field of amplitude $\Phi_0$. 
To determine the $\Phi_0$-dependent part of $\Omega_0$ we calculate 
\begin{equation}\label{eq:DefGamma}
\Gamma(\Phi_0) \equiv 2 \frac{\partial \Omega_0}{\partial \Phi_0} = 
\sum_\alpha\left[f\left(\xi_\alpha + \frac{\Phi_0}{2}\right) - f\left(\xi_\alpha - \frac{\Phi_0}{2}\right)\right]\ ,
\end{equation}
where $\xi_\alpha \equiv \epsilon_\alpha - \mu$ and $f(\epsilon) \equiv  (\exp[\beta \epsilon]+1)^{-1}$ is the Fermi distribution function.
At zero temperature $\Gamma(\Phi_0)$ is the number of orbital levels between $\mu - \Phi_0/2$ and $\mu + \Phi_0/2$. Assuming a constant density of states (equidistant spectrum) we obtain 
\begin{equation}\label{eq:GammanuPhi}
\Gamma = \nu \Phi_0\ ,\quad
\frac{\partial \Omega_0}{\partial \Phi_0} = - \frac{1}{2}\,\nu \Phi_0\ ,\quad \Omega_0 = const. - \frac{\nu\Phi_0^2}{4}\ .
\end{equation}
Strictly speaking (\ref{eq:GammanuPhi}) is valid at temperatures higher than the level spacing $\delta$, 
i.e., for $T \gg \delta = \nu^{-1}$. At lower temperatures step-like dependencies are expected. 
Yet, in a "coarse-grained" sense (\ref{eq:GammanuPhi}) holds at lower temperatures as well.
Finally, 
\begin{eqnarray}
{\cal S}_{\Phi} ^{(0)}=const.  - \frac{\beta \Phi_0^2}{4J^\ast} - \sum_{m\neq 0} \frac{\beta}{4J}\, \delta\Phi_m \delta\Phi_{-m}\ ,
\label{eq:S0a}
\end{eqnarray}
where $1/J^\ast \equiv 1/J - \nu$. As we consider the regime close to Stoner instability, we 
have $J^\ast \gg J > 0$. Realistically, the quantum dots are disordered and the assumption of an 
equidistant spectrum is too naive~\cite{BurmistrovDisorder}. Due to disorder we should have 
\begin{equation}\label{eq:DisorderOmega}
\Omega_0 = const. - \frac{\bar\nu\Phi_0^2}{4} + \delta\Omega_0\ ,
\end{equation}
where $\bar\nu$ is the average density of states. This question was originally addressed 
by Kurland et al.~\cite{PhysRevB.62.14886} and was recently analyzed by Burmistrov et al.~\cite{BurmistrovDisorder}. 
Roughly, $\delta \Omega_0$ is of order $\pm O(1) \Phi_0$. In the present paper we disregard disorder. 

\subsection{The first order contribution ${\cal S}_{\Phi} ^{(1)}$}
\label{sec:action-s1}
In the first order in $\eta$ (Eq.~\ref{Inv7}) we obtain
\begin{eqnarray}
\label{Inv8}
&&{\cal S}_{\Phi} ^{(1)}=- \frac{1}{\beta} \int\limits_0^\beta d\tau\, \mathrm{tr} \left[G^0(\tau,\tau)\eta(\tau)\right]= - \sum_n \mathrm{tr}\left[G^0(\varepsilon_n)\frac{1}{\beta}\int\limits_0^\beta d \tau\, \eta(\tau)\right]\ ,
\end{eqnarray}
where $G_{\alpha,\alpha'}^0(\varepsilon_n) \equiv  \delta_{\alpha,\alpha'}\left(i \varepsilon_n -
\xi_{\alpha} - \frac{\sigma_z}{2}\, \Phi_0\right)^{-1}$ and $\varepsilon_n = \pi(2n+1)/\beta$ are the fermionic Matsubara frequencies. Calculating the sum over $\varepsilon_n$ we obtain 
\begin{eqnarray}
\label{Inv9}
&&{\cal S}_{\Phi} ^{(1)}=-{1\over 2}\sum_{\alpha}\mathrm{tr}\left(\left[
\left(f\left(\xi_{\alpha} +\frac{\Phi_0}{2}\right) + f\left(\xi_{\alpha} -\frac{\Phi_0}{2}\right) -1 \right)\sigma_0  +
\left(f\left(\xi_{\alpha} +\frac{\Phi_0}{2}\right) - f\left(\xi_{\alpha} -\frac{\Phi_0}{2}\right) \right)\sigma_z\right]
\int\limits_0^\beta d \tau\, \eta(\tau)\right)\ .
\end{eqnarray}
From $\mathrm{tr}(\eta(\tau))=0$ we conclude that
\begin{eqnarray}
\label{Inv10}
&&{\cal S}_{\Phi} ^{(1)}={\Gamma\over 2} \int_{0}^{\beta}
d\tau\, \mathrm{tr}\left(\sigma_z \eta \right)={i\Gamma\over 2} \int_{0}^{\beta}
d\tau\,\dot \phi \,(1-\cos\theta)\ ,
\end{eqnarray}
where $\Gamma$ was defined in Eq.~(\ref{eq:DefGamma}).
The contribution to the effective action, given by Eq.\ref{Inv10}, is proportional to the Berry phase. 
The coefficient in front of the Berry phase $\Gamma(\Phi_0)$ is roughly equal to the the number of single occupied levels, 
i.e., the number of uncompensated spins which acquire the Berry phase. 

\subsection{The second order contribution ${\cal S}_{\Phi}^{(2)}$}
\label{sec:action-s2}
To calculate the second order contribution we introduce the notations
$G^0_{\uparrow/\downarrow} \equiv (i \varepsilon_n - \xi_{\alpha} \mp \Phi_0/2)^{-1}$ and $\sigma_{\uparrow/\downarrow}
\equiv{1\over 2} (1 \pm \sigma_z)$ and we obtain 
\begin{eqnarray}
\label{Inv11}
&&{\cal S}_{\Phi}^{(2)}=-{1\over 2}\sum_{\alpha}\sum_{n,m} \mathrm{tr}\left(\eta_{m} G^0_{n+m} \eta_{-m}
G^0_{n}\right)=-{1\over 2}\sum_{\alpha}\sum_{n,m} \mathrm{tr}\left(\eta_{m} (G^0_\uparrow
\sigma_\uparrow + G^0_\downarrow\sigma_\downarrow)_{n+m} \eta_{-m}
(G^0_\uparrow\sigma_\uparrow + G^0_\downarrow\sigma_\downarrow)_{n}\right)\ .
\end{eqnarray}
Here $\eta_m \equiv \frac{1}{\beta}\int\limits_0^{\beta}d\tau\, \eta(\tau)e^{i\omega_m \tau}$ and 
$\omega_m =2\pi m/\beta$ are bosonic Matsubara frequencies. For the transverse, i.e., the spin flipping part of $\eta$, we obtain
\begin{eqnarray}
\label{Inv13}
{\cal S}_{\Phi,\perp}^{(2)}= {\beta\Gamma\over 2}\sum_{m} \left[\mathrm{tr}\left(\eta_m \sigma_\uparrow \eta_{-m}
\sigma_\downarrow\right){1\over \Phi_0 - i \omega_m} +  
\mathrm{tr}\left(\eta_{m} \sigma_\downarrow \eta_{-m}
\sigma_\uparrow\right){1\over \Phi_0 + i\omega_m}\right] \ .
\end{eqnarray}
In the adiabatic limit $|\omega_m| \ll  \Phi_0$ this gives
\begin{eqnarray}
\label{Inv13perp}
{\cal S}_{\Phi,\perp}^{(2)} \approx {\beta \Gamma\over \Phi_0} \sum_{m} \mathrm{tr}\left(\eta_m \sigma_\uparrow \eta_{-m}
\sigma_\downarrow\right)= {\Gamma\over \Phi_0}\int d\tau\, \mathrm{tr}\left(\eta(\tau) \sigma_\uparrow \eta(\tau)
\sigma_\downarrow\right)\ .
\end{eqnarray}
Hence, substitution of Eq.\ref{Inv7} leads to the result
\begin{eqnarray}
&&{\cal S}_{\Phi,\perp}^{(2)} =-{\Gamma\over 4\Phi_0}\int\limits_0^\beta d\tau \left[\dot \theta^2 + (\sin
\theta\, \dot \phi)^2\right]=- {\Gamma\over 4\Phi_0}\int\limits_0^\beta d\tau \,\dot{{\vec n}}^2.
\label{Inv15}
\end{eqnarray}
where $\dot{{\vec n}}$ is the angular velocity describing the motion of the 
magnetization vector $\vec{\Phi}$ on the sphere. 

For the longitudinal part of $\eta$ we obtain
\begin{eqnarray}
\label{Inv13par}
{\cal S}_{\Phi,\parallel}^{(2)}= {\beta\nu\over 2} \left[\mathrm{tr}\left(\eta_{m=0} \sigma_\uparrow \eta_{m=0}
\sigma_\uparrow\right) + \mathrm{tr}\left(\eta_{m=0} \sigma_\downarrow \eta_{m=0}
\sigma_\downarrow\right)  \right] = -\frac{\nu}{4\beta}\left(\int_{0}^{\beta}
d\tau\,\dot \phi \,(1-\cos\theta)\right)^2\ .
\end{eqnarray}
Finally ${\cal S}_{\Phi}^{(2)} = {\cal S}_{\Phi,\perp}^{(2)} + {\cal S}_{\Phi,\parallel}^{(2)}$. 
Both this terms are $\sim$ $\beta \nu (\dot{\vec n})^2$. In comparison 
${\cal S}_{\Phi}^{(1)}$ of (\ref{Inv10}) is $\sim$ $\beta \nu \Phi_0 \dot{\vec n}$. 
Thus for all adiabatic frequencies $|\omega_m| \ll \Phi_0$ the Berry's phase action ${\cal S}_{\Phi}^{(1)}$ dominates. 
In what follows we restrict ourselves to the first order, Berry phase, contribution ${\cal S}_{\Phi}^{(1)}$.

\section{Partition function and magnetic susceptibility}
\label{secIV}
\noindent
{\it{In this section, after performing the path integration over the adiabatic paths of $\vec n(\tau)$, 
we obtain the partition function of the problem as a function of $\Phi_0$ only. This allows 
us to calculate the magnetic susceptibility consisting of the Pauli and Curie terms.}}\\

Above we have obtained the following effective action for the adiabatic paths
\begin{eqnarray}
\label{Inv18}
{\cal S}_\Phi \approx
const.  - \frac{\beta \Phi_0^2}{4J^\ast} - \sum_{m\neq 0} \frac{\beta}{4J}\, \delta\Phi_m \delta\Phi_{-m}
+ {i\Gamma\over 2} 
\int\limits_0^\beta d\tau \,(1-\cos\theta) \dot \phi \ .
\end{eqnarray}
This action governs fluctuations at low frequencies $|\omega_m|<\Phi_0$. Thus, at these frequencies the system 
reduces to a large spin of amplitude $\sim \Gamma$.  Indeed the last term of the action (\ref{Inv18}) is just the well known 
Wess-Zumino action of a free spin. While a partition function of a true free spin is trivial to calculate, our spin "lives" only 
at the adiabatic frequencies. We, thus, define our functional integral for the partition function as an integral over 
all paths whose frequency scale is cut off by $\Phi_0$. We have
\begin{equation}
\label{Inv19}
{\mathcal {Z}}=\mathcal{N} \int \mathcal{D}^3 \Phi \, e^{{\cal S}_\Phi}\quad, {\rm where}\quad
\mathcal{D}^3 \Phi \equiv \prod_i d^3\Phi_i\ .
\end{equation}
In Eq.\ref{Inv19} the index $i$ defines a lattice partition of the interval $[0,\beta]$
defined so as to limit frequencies to values $|\omega_m|<\Phi_0$. The measure $d^3\Phi_i$ is an ordinary 
Cartesian measure. The Hubbard-Stratonovich transformation and integrating out the fermions
produce the following normalization factor
\begin{eqnarray}
&&{\mathcal{N}}= \tilde{\mathcal{N}}\left({\beta\over J}\right)^{3\left(N+{1\over 2}\right)}
\label{Inv21}
\end{eqnarray}
Here $\tilde{\mathcal{N}}=1/\left[\pi^{\left(N + {1\over 2}\right)}4^{3\left(N+{1\over 2}\right)}\right]$
and $N=\Phi_0/2\pi T \gg 1$ is the number of positive Matsubara frequencies taken into account (those that are 
below the cut off $\Phi_0$).

\subsection{Calculating the path integral}
\label{sec:doing-integral-at}

We are guided by the idea that the most important paths are those of almost constant radius 
$\Phi(\tau) \approx \Phi_0$. Indeed, as can be seen from (\ref{Inv18}) the longitudinal fluctuations 
$\delta \Phi$ are suppressed by a factor containing the bare $J$. By contrast, close to the 
Stoner instability, the zero frequency amplitude $\Phi_0$ is only weakly suppressed ($J^\ast \gg J$). The larger 
is $\Phi_0$, the bigger is the phase volume of possible transverse fluctuations. The latter are only ``penalized'' 
by the Berry phase term. An entropic ``phase space'' argument, outlined above, reveals
that $\Phi_0$ should assume a large value. Below we find out that the typical value of $\Phi_0$ is of the order
of $J^\ast$, whereas we have $|\delta \Phi_m | \sim \sqrt{TJ}$ and $|\delta\Phi(\tau)| \sim \sqrt{\sum_m |\delta\Phi_m|^2} 
\sim \sqrt{J^\ast J} \ll \Phi_0$. To derive all these, we begin by converting the measure to a polar one, which is then 
adjusted to an integration over paths of almost constant radius $\Phi_0$:
\begin{eqnarray}
&&\mathcal{D}^3 \Phi=\prod_{i \in time~intervals} \Phi_i^2 d\Phi_i d \vec n_i =e^{2\sum_i \ln
\Phi_i}\prod_i  d\Phi_i d \vec n_i \simeq \Phi_0^{2\left(2N+1\right)} e^{-{1\over
\Phi_0^2}\sum_i \delta \Phi_i^2}  \prod_i  d\Phi_i d\vec n_i \nonumber \\
&&\simeq
\Phi_0^{4 N} \Phi_0^2 d\Phi_0 \prod_m  d \delta \Phi_m \mathcal{D}\vec n\ ,
\label{Inv22}
\end{eqnarray}
where in the last identity we have dropped the term in the exponent
\begin{equation}
{1\over\Phi_0^2 }\sum_i \delta\Phi_i^2  \approx {1\over{\Phi_0^2 \Delta \tau}} \int\limits_0^\beta d\tau\, (\delta\Phi(\tau) )^2
\simeq  {\beta \omega_c\over{2\pi \Phi_0^2}} \sum_{|\omega_m| < \omega_c}  |\delta \Phi_m|^2\ .
\end{equation}
Here $\omega_c\sim 1/\Delta\tau$ is the ultra-violet cutoff. Taking consistently the adiabatic cutoff 
$\omega_c \sim \Phi_0$ we observe that this term  
is negligible in comparison with the $\sim \beta/J$ term in the action (\ref{Inv18}).

We now perform the Gaussian integration over the longitudinal fluctuations $\delta \Phi_m$, the net effect of
which is the partial cancellation of the normalization factor (\ref{Inv21}) by $(\beta/J)^N$. We are left with 
\begin{eqnarray}
{\cal {Z}}=\tilde{\mathcal{N}}\left({\pi\over 2}\right)^{N}\left({\beta\over J}\right)^{3\over 2}\int\Phi_0^2 d\Phi_0
\left({\Phi_0^2\beta\over J}\right)^{2N} \exp\left[{-{\beta\over 4J^\ast} \Phi_0^2}\right]\,\int \mathcal{D}\vec n 
\,\exp\left[ {i\Gamma\over 2} \int\limits_0^{\beta} d\tau \,\dot \phi\,(1-\cos\theta) \right]\ .
\label{Inv23}
\end{eqnarray}
We next proceed to the integral over the transverse fluctuations, \ie, the fluctuations of $\vec n(\tau)$. 
These fluctuations are ``penalized'' by the Berry phase term in the action. The Berry phase term is given 
by the solid angle swept by the ${\vec n(\tau)}$ path. 
Since $\Gamma \sim\nu \Phi_0\gg 1$ (this has to be checked for self consistently),
these solid angles must remain small, and we can restrict ourself to a Gaussian expansion around a static value of 
${\vec n}$, namely ${\vec n}_0$, the relevant paths will be given by the 
small variations around this static ${\vec n}_0$. At the end we should average over all possible directions 
of ${\vec n}_0$.

The Gaussian integration is easily performed if we note that the integration measure $\mathcal{D}\vec n$
is given by $\prod_i d \phi_i d (1-\cos\theta_i)$. Thus, introducing $y \equiv 1-\cos\theta$ we obtain 
\begin{equation}
\int \mathcal{D}\vec n 
\,\exp\left[ {i\Gamma\over 2} \int\limits_0^{\beta} d\tau \,\dot \phi\,(1-\cos\theta) \right] \approx \int \mathcal{D} \phi \mathcal{D} y \, \exp\left[\frac{i\Gamma}{2}\, \int\limits_0^{\beta}\,d\tau\, \dot\phi y\right] \approx \prod_{m=1}^N \left({1\over \beta\Gamma\omega_m}\right)^2
\ .
\end{equation} 
In evaluation of the integral we extended the integration limits of both $\phi$ and $y$ to $[-\infty,\infty]$ 
even though, e.g., $y \in [0,2]$. This is justified for $\Gamma\gg 1$.
Thus we obtain
\begin{align}
&&{\cal{Z}}=\tilde{\mathcal{N}}\left({\pi\over 2}\right)^{N}
\left({\beta\over J}\right)^{3\over 2}\int\Phi_0^2 d\Phi_0
\left({\Phi_0^2\beta\over J}\right)^{2N}e^{-{\beta\over 4J^\ast}
\Phi_0^2}\prod_{m=1}^N \left({1\over \beta\Gamma\omega_m}\right)^2 
=\tilde{\mathcal{N}}\left({\pi\over 2}\right)^{N}
\left({\beta\over J}\right)^{3\over 2}\left(\frac{1}{J\nu}
\right)^{2N}\int\Phi_0^2 d\Phi_0e^{-{\beta\over 4J^\ast}\Phi_0^2}
\prod_{m=1}^N \left(\Phi_0\over \omega_m\right)^2\ .
\label{Inv24}
\end{align}
The product appearing in Eq.~\ref{Inv24} is cutoff at $m=N$, \ie, at $\omega_m = \Phi_0$. 
This hard cutoff is an artifact of our rather hand-waving approach. Within this method 
we have no way to determine what kind of cut off should be employed. One possibility, which is 
supported by the calculation in Cartesian coordinates provided in \ref{appendixA}, is to use a soft cut off. 
This gives
\begin{eqnarray}
&&\prod_{m=1}^N  \left(\Phi_0\over \omega_m\right)^2  \approx  \prod_{m=1}^\infty  \left(\Phi_0^2 + \omega_m^2\over \omega_m^2\right)  =  \frac{\sinh\left[\frac{\beta\Phi_0}{2}\right]}{\frac{\beta\Phi_0}{2}} 
\simeq \frac{\exp\left[\frac{\beta\Phi_0}{2}\right]}{\beta\Phi_0} \ . 
\label{Inv25}
\end{eqnarray}
Evidently, employing a different cut off, one would obtain a result which is different in the 
exponential. For example, the hard cut off gives $\frac{\beta\Phi_0}{\pi}$ in the exponent. This would slightly 
modify the numerical coefficients in the final result for the susceptibility. We keep here the 
soft cut off result, as it is supported by the calculation in \ref{appendixA}, and provides an excellent 
approximation to the exact solution~\cite{PismaJETP.92.202}.

We approximate the $(1/\nu J)^N$ factor in (\ref{Inv24}) by 1, which is 
legitimate near the Stoner transition. Finally, integrating over $\Phi_{0}$ we obtain
\begin{eqnarray}
\label{PFintPhia}
&&{\mathcal{Z}} =\tilde{\mathcal{N}}\left({\pi\over 2}\right)^{N}
\left(\frac{\beta}{J}\right)^{3/2} \int\limits_0^\infty d\Phi_0  4\pi\Phi_0^2 
\exp{\left[-\frac{\beta \Phi_0^2}{4J^\ast}\right]}\cdot
\frac{\sinh\left[\frac{\beta\Phi_0}{2}\right]}{\frac{\beta\Phi_0}{2}}
= 4\pi\tilde{\mathcal{N}}\left({\pi\over 2}\right)^{N}
\left(\frac{J^\ast}{J}\right)^{3/2}\,\exp\left[{\frac{\beta J^\ast}{4}}\right]\ . 
\end{eqnarray}

\subsection{Magnetic susceptibility}
\label{sec:mag-sus}

We obtain the magnetic susceptibility using Eq.\ref{PFintPhia} as follows
\begin{eqnarray}
&& \chi= \frac{1}{3}\,\frac{\partial \ln {\mathcal {Z}}}{\partial J} = \frac{1}{2}\,\frac{\nu}{(1-J\nu)} + \frac{\beta}{12}\,\frac{1}{(1-\nu J)^2}\ .
\label{Inv29}
\end{eqnarray}
We observe that it consists of a Pauli and a Curie contribution. 
In comparison the exact solution of Ref.~\cite{PismaJETP.92.202} reads
\begin{eqnarray}
{\cal {Z}}_{exact} \simeq \left(\frac{J^\ast}{J}\right)^{3/2}\,\exp\left[{\frac{\beta (J^\ast - J)}{4}}\right]\ .
\end{eqnarray}
and the magnetic susceptibility
\begin{eqnarray}
\chi_{exact}  = \frac{1}{2}\,\frac{\nu}{(1-J\nu)} + \frac{\beta}{12}\,\left[\frac{1}{(1-\nu J)^2}-1\right]\ .
\label{chiexact}
\end{eqnarray}
In close vicinity to the Stoner instability, \ie, $\delta\sim\nu^{-1}\sim J$ the extra $-1$ factor in $\chi_{exact}$
is immaterial and we obtain an extremely good approximation to the exact result.  

The Pauli-like (with an upward renormalized g-factor) susceptibility (first term in Eq.~(\ref{Inv29}))
dominates when $T  \gg J^\ast$. In the low temperature regime, $T \ll J^\ast$, the Curie-like part 
(second term) dominates. In this regime the average spins scales as $\sqrt{<{\bf {S}}>^{2}}\sim J^\ast/\delta$.
This Curie like contribution in the magnetic response can be tested in materials close to the Stoner instability 
such as Pd ($J/\delta = 0.83$) or ${\rm YFe_{2}Zn_{20}}$ ($J/\delta = 0.94$)~\cite{PCCanfield}. It is important 
to understand that the Curie part of the susceptibility represents a mesoscopic effect. The density of states 
of a \qdd $\nu$ scales linearly with the volume of the dot, $\nu\sim V$. On the other hand $J \sim 1/V$.
Hence the Pauli like part of the magnetic susceptibility is proportional
to the volume $V$ and is an extensive quantity. On the other hand, the Curie susceptibility is 
intensive as $\nu J$ is scale invariant. Therefore for a fixed temperature $T$, if one gradually 
increases the size of the system, the Pauli part grows and, eventually, 
the Curie susceptibility becomes negligible compared to the Pauli one. 
\begin{figure}
\hskip -1.2cm
\includegraphics[width=10.0cm,height=8.0cm]{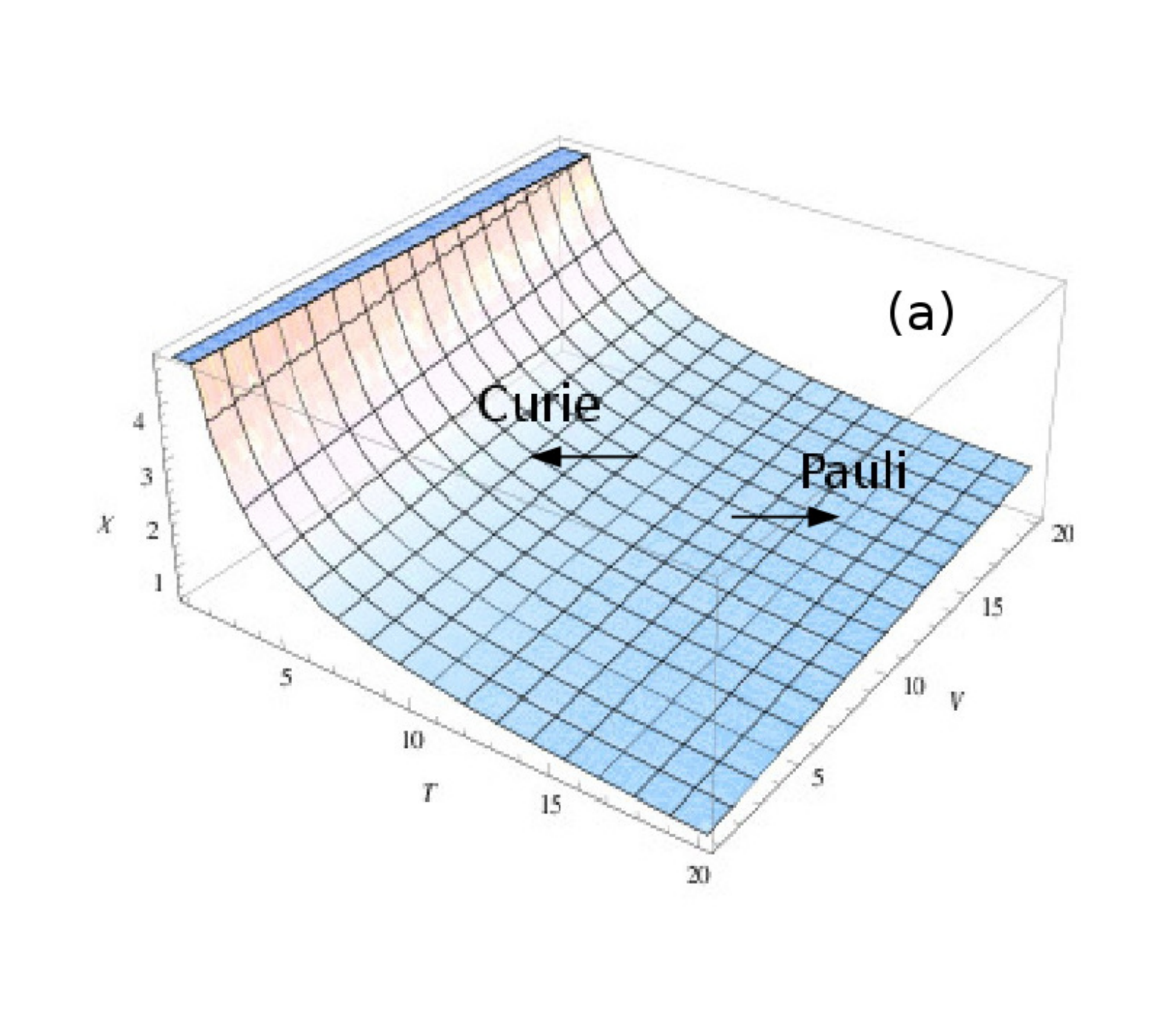}
\hskip 0.5cm
\includegraphics[width=7.0cm,height=6.0cm]{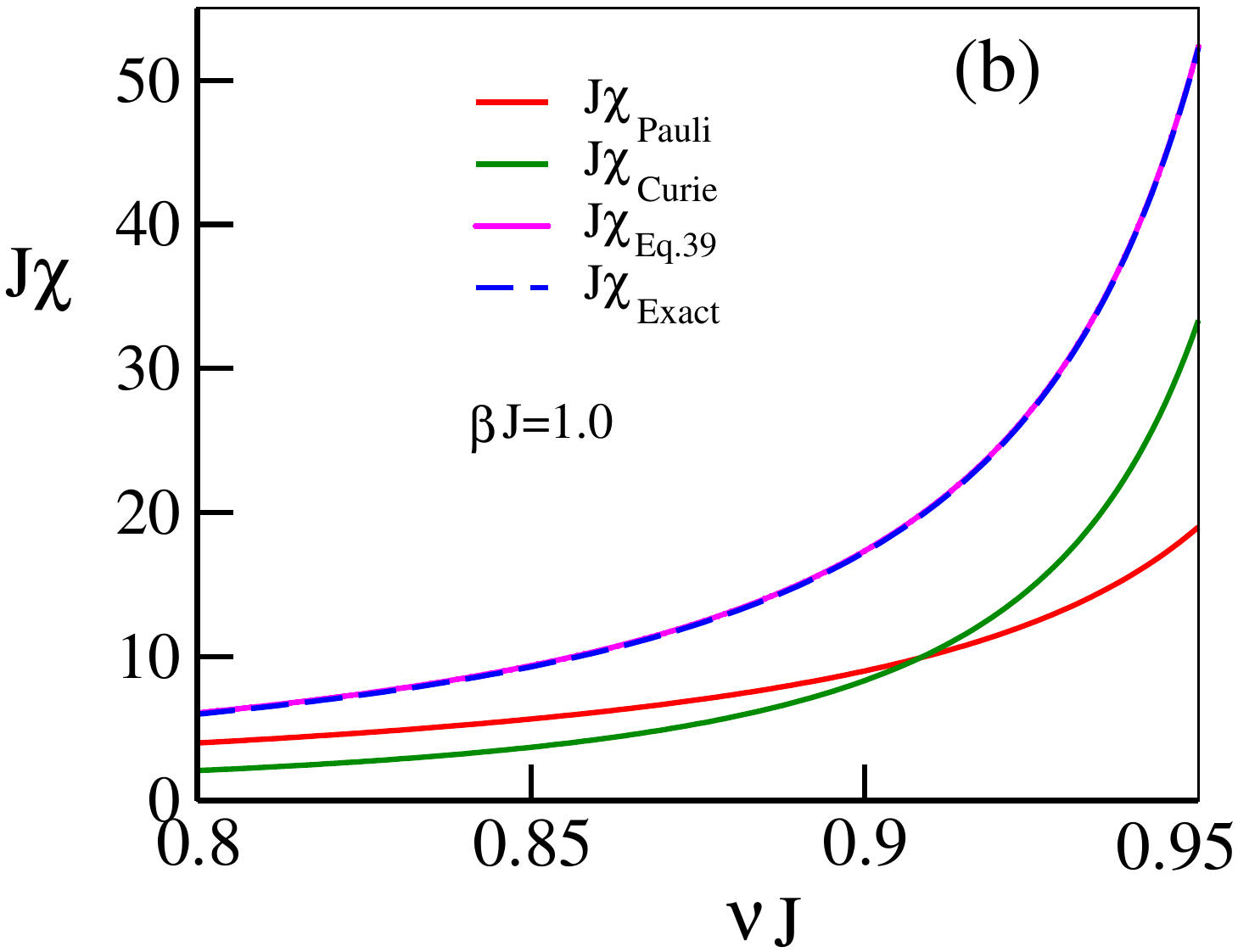}
\caption{(Color online) (a)~Magnetic susceptibility of an isolated \qdd in the $T$ $-$ $V$ plane where
$T$ is the temperature and $V$ is the volume of the background $3D$ material. Here $\nu J=0.9$
\ie~close to the Stoner instability point. (b)~Dimensionless susceptibility $J \chi$
is shown as a function of the Stoner parameter $\nu J$. Here the dimensionless parameter $\beta J=1.0$.
The red, green, magenta and blue curve correspond to the Pauli like susceptibility (first term in Eq.~(\ref{Inv29})),
Curie like susceptibility (second term in Eq.~(\ref{Inv29})), our approximate expression for the magnetic 
susceptibility (Eq.~(\ref{Inv29})) and the exact expression of susceptibility (Eq.~(\ref{chiexact})) respectively.}
\label{figsusceptibility}
\end{figure}
This behavior is shown in Fig.\ref{figsusceptibility}(a) where $\nu J$ is chosen to be in close vicinity of
the Stoner instability regime. To illustrate the nature of Pauli and Curie like 
suceptibility more, we show their behavior as a function of the Stoner parameter $\nu J$ in 
Fig.~\ref{figsusceptibility}(b). It is clear that for a fixed dimensionless parameter $\beta J$, the 
Curie susceptibility (green curve) becomes more dominant over the Pauli one (red curve) as we approach towards 
the Stoner instability point. On the other hand, the blue curve shows the behavior of $J \chi_{exact}$ which
in close vicinity to the Stoner instability point matches quite well to our approximate answear (Eq.~(\ref{Inv29})),
shown by the magenta curve in Fig.~\ref{figsusceptibility}(b).

\section{Summary and Discussion}
\label{secV}
In this paper we have considered an isolated \qdd with an
isotropic exchange interaction. In the path integral \hs\  formulation this problem leads to an 
effective non-abelian action. Here we have presented an approximative geometric approach
in which the Berry phase controls the dynamics of the direction $\vec n$ of the \hs\ 
magnetization vector $\vec \Phi$. Close to the Stoner limit, i.e., for $\nu J \rightarrow 1$,
our approach reproduces well the exact solution of Ref.~\cite{PismaJETP.92.202} which comes 
at the expense of a hard calculation based on the generalized Wei-Norman-Kolokolov method~\cite{IVKolokolov}.
Note that even zero frequency observables (for \eg~susceptibility), involve summation over finite frequency
fluctuations (Eq.~(\ref{effS1})), including both low and high frequency contributions.
Although the exact result describes both the low and high energy contents of susceptibility, we here show that 
if one is interested in the low energy regime of long range fluctuations, the information of the
exact result can be obtained from our physically motivated and user friendly approach of the invariant action.
However if one insists on knowing ultraviolett behavior, then one may resort to an equally simple gaussian
expansion around stationary points. Hence, our approximative geometric approach covers much of the contents of 
the observable, in a manner suitable for further generalization. Therefore, we believe that this approach could 
be very useful in situations in which the exact method is inapplicable. These may be, e.g., the problems of charge 
and spin transport via quantum dots coupled to normal or ferromagnetic leads.

Strictly speaking our results are valid for $\delta = \nu^{-1} \sim J < T < J^{\ast}$. Yet, for the 
equidistant spectrum assumed in this paper we do not expect major changes at lower temperatures.  
In the presence of disorder the low temperature regime $T < J^{\ast}$ is itself more 
subtle~\cite{BurmistrovDisorder}.

Finally, it is important to mention that, as discussed around Eq.~(\ref{eq:DisorderOmega}), disorder 
can influence the grand canonical potential of non-interacting electrons $\Omega_0$ quite essentially. 
Yet, it is easy to show that the Berry phase part of the action is quite insensitive to disorder.

\section{Acknowledgments}

This work was supported by GIF, Einstein Minerva Center, Sonderforschungsbereich TR 12 of the Deutsche 
Forschungsgemeinschaft, EU FP7 grant GEOMDISS, the Russian-Israel scientific research cooperation 
(RFBR Grant No. 11-02-92470 and IMOST  3-8364), the Council for Grant of the President of 
Russian Federation (Grant No. MK-296.2011.2), RAS Programs ``Quantum Physics of Condensed Matter'' 
and ``Fundamentals of nanotechnology and nanomaterials'', the Russian Ministry of Education and Science under 
contract No. P926.

We acknowledge useful discussions with  
Gabriele Campagnano, Igor Lerner, Mikhail Kiselev, J\"{u}rgen K\"{o}nig and Alexander Mirlin. We are grateful 
to Ganpathy Murthy for providing us with notes of his calculations on ''Universal interacting crossover 
regime in two-dimensional quantum dots`` (Ref.\cite{PhysRevB.77.073309}) and a detailed explanation. 

\appendix
\section{Unimportance of longitudinal fluctuations}
\label{appendixAA}
\noindent
{\it{Here we show why the longitudinal fluctuations of $\delta\vec{\Phi}$ can be disregarded.}}

Following the spirit of Ref.~\cite{KamenevGefen} we gauge out in Eq.~(\ref{Inv6}) the fluctuations $\delta \Phi$ by using a 
(non-unitary) transformation $V=\exp\left[-\frac{\xi}{2}\,\sigma_z\right]$, where $\xi(\tau)\equiv \int\limits_0^\tau\,\delta\Phi(\tau')\,d\tau'$. We obtain
\begin{eqnarray}
\sum_{\alpha}\mathrm{tr\;ln}\left(-\partial_\tau - \epsilon_{\alpha} + \mu - \Phi(\tau)\,\frac{\sigma_z}{2} -
R^{-1}\partial_\tau R\right) =\sum_{\alpha}\mathrm{tr\;ln}\left(-\partial_\tau - \epsilon_{\alpha} + \mu - \Phi_0\,\frac{\sigma_z}{2} -
\eta(\tau) \right)\ ,
\label{Inv6aa}
\end{eqnarray}
where 
\begin{eqnarray}
\eta(\tau)&\equiv& V^{-1} R^{-1}\left(\partial_\tau R \right)V  \nonumber\\
&=&-{i\over 2} \dot \phi (\cos \theta-1) \sigma_z +{i\over 2}\,\exp\left[\frac{\xi}{2}\,\sigma_z\right] \left(\dot \phi \left[\sin\theta (\cos\phi\sigma_x+\sin\phi\sigma_y)\right] 
- \dot \theta \left[\cos\phi \sigma_y - \sin\phi \sigma_x\right]\right)\, \exp\left[-\frac{\xi}{2}\,\sigma_z\right]  \  .\nonumber\\
\label{Inv7aa}
\end{eqnarray}
This means that, in fact, $\eta(\tau)$ appearing in (\ref{Inv7}) is the expression (\ref{Inv7aa}). 
We observe that the longitudinal part of $\eta$ of (\ref{Inv7aa}), which is responsible for the Berry phase term in the 
action, does not contain the operators $V, V^{-1}$ and, thus, is not influenced by the longitudinal fluctuations 
$\delta \Phi$ which appear only in the factor $\xi$. Moreover, substituting $\eta$ of (\ref{Inv7aa}) 
into Eq.~(\ref{Inv13perp}) we observe that in the adiabatic limit the longitudinal fluctuations drop out also in the 
second order terms. Thus, disregarding the longitudinal fluctuations in (\ref{Inv6a}), and the factor $V$ 
altogether, was justified.

\section{Expansion in small transverse fluctuations}
\label{appendixA}
\noindent
{\it{Here we present an alternative method of calculating the path integral 
in Cartesian coordinates. The advantage is the higher level of accuracy in handling the 
integration measure. An adiabatic cut off need not be postulated here. 
Rather, a soft adiabatic cut off appears naturally.}}

\subsection{Effective action}
\label{appendixA1}

\begin{figure}
\begin{center}
\includegraphics[width=6.0cm,height=8.0cm]{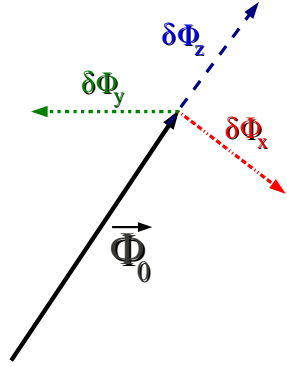}
\caption{(Color online) Schematic of the auxiliary bosonic \hsd field $\vec{\Phi}$ where 
$\vec{\Phi}_{0}$ is the zero-component and $\delta\Phi_{x},\delta\Phi_{y},\delta\Phi_{z}$ are the 
non-zero Matsubara components. Among the three non-zero Matsubara components $\delta\Phi_{z}$ 
is the longitudinal and $\delta\Phi_{x},\delta\Phi_{y}$ are the transverse fluctuations respectively. 
Here in our analysis the longitudinal component $\delta\Phi_{z}$ is chosen in accordance with the 
direction of $\vec{\Phi}_0$ and $\delta\Phi_{x},\delta\Phi_{y}$ are the two transverse componets.}
\label{figfluctuations}
\end{center}
\end{figure}
We start from the Hamiltonian (\ref{eq:HJ}) and rewrite the action (\ref{Inv1}) 
in the Matsubara representation as
\begin{eqnarray}
&&{\cal S}_\Psi = \beta  \Bigg[\sum_{\alpha\sigma,n} \bar\psi_{\alpha\sigma,n}
(i\varepsilon_n -\epsilon_\alpha +\mu) \psi_{\alpha\sigma,n} 
+ J  \sum_m \left[\sum_{\alpha\sigma_x \sigma_y} \bar \psi_{\alpha\sigma_x} 
{\bf S}_{\sigma_x \sigma_y} \psi_{\alpha\sigma_y} \right]_{m}  
\left[\sum_{\alpha\sigma_x \sigma_y} \bar \psi_{\alpha\sigma_x} 
{\bf S}_{\sigma_x \sigma_y} \psi_{\alpha\sigma_y} \right]_{-m}\Bigg] \ .
\label{Smatsu}
\end{eqnarray}
We apply the \hsd transformation on Eq.\ref{Smatsu} to obtain an effective action quadratic in the fermionic 
fields, with an auxiliary vector bosonic field ($\vec{\Phi}$) for the spin degrees of freedom. Hence, the effective
action reads
\begin{align}
&&{\cal S}_{\Psi,\Phi} = \beta\Bigg[\sum_{\alpha\sigma,n} \bar\psi_{\alpha\sigma,n}
(i\varepsilon_n -\epsilon_\alpha +\mu) \psi_{\alpha\sigma,n} - \sum_m \vec  \Phi_m \cdot
\left[\sum_{\alpha\sigma_1 \sigma_2} \bar \psi_{\alpha\sigma_1} {\bf S}_{\sigma_1 \sigma_2} 
\psi_{\alpha\sigma_2} \right]_{-m} 
- \sum_{m} \frac{ \vec \Phi_m \vec \Phi_{-m}}{4J} \Bigg] \ , 
\label{SHS}
\end{align}
where the Matsubara expansion for the bosonic \hsd real vector field $\vec \Phi$ reads
\begin{equation}
\vec \Phi(\tau) =\sum_{m} \vec \Phi_m e^{-i\omega_m \tau}  = \vec \Phi_0 + \sum_{m\neq 0} \delta \vec \Phi_m e^{-i\omega_m \tau} \ .
\end{equation}
Here $\vec \Phi_0$ is the zero component of the bosonic \hsd real vector field $\vec \Phi$ and
$\delta \vec \Phi_m = \vec \Phi_m$ are the non-zero ones. 

Our strategy is to, first, integrate over the non-zero Matsubara components $\delta \vec \Phi_m$ while 
keeping $\vec \Phi_0$ fixed and, then, integrate over all possible $\vec \Phi_0$. The vectors 
$\delta \vec \Phi_m$ have three components among which 
one is longitudinal and the other two are the transverse with respect to the current direction of 
$\vec \Phi_0$, which is schematically shown
in Fig.\ref{figfluctuations}. Moreover we choose the basis for $\delta \vec \Phi_m$ 
in accordance with the direction of $\vec \Phi_0$, \ie, the axis $z$ for $\delta \vec \Phi_m$ is along 
$\vec \Phi_0$. In other words for the time being we break the symmetry of the isotropic problem choosing 
a particular direction of $\vec \Phi_0$ and select the basis for $\delta \vec \Phi_m$ 
in accordance with the current direction of $\vec \Phi_0$. Note that at the end of the day
one should integrate over all $\vec \Phi_0$ in order to restore the global symmetry 
of the problem and obtain an expression for the partition function and the susceptibility
compatible with Eqs.\ref{PFintPhia} and \ref{Inv29}. In terms of $\Phi_{0}$ and the cartesian
fluctuations around it, Eq.\ref{SHS} can be rewritten as 
\begin{eqnarray}
{\cal S}_{\Psi,\Phi} = \beta \left[{\rm Tr_{\rm spin}}\sum_{\alpha,n_1,n_2} \bar\psi_{\alpha,n_1}
\left[\left(G_{0,\alpha,n_1}^{-1}\right)\delta_{n_1,n_2} 
- \delta \vec \Phi_{(n_1-n_2)} \cdot {\bf S}  \right] \psi_{\alpha,n_2} -\frac{\Phi_0^2}{4J} 
- \sum_{m\neq 0} \frac{\delta \vec \Phi_m \delta \vec \Phi_{-m}}{4J} \right]\ , 
\end{eqnarray}
where $G_{0,\alpha,n}^{-1} = i\varepsilon_n -(\epsilon_\alpha - \mu) - \Phi_0 S_z$ is the
single particle Green's function of the electrons subject to a constant magnetic field $\Phi_0$.
For the partition function we obtain
\begin{eqnarray}
\label{Z_HS}
\mathcal{Z}(\mu) =\left(\frac{1}{4\pi}\right)^{3\left(N + {1\over 2}\right)}
\left(\frac{\beta}{J}\right)^{3/2}
\left(\frac{\beta}{J}\right)^{3N} 
\int D\bar\Psi D\Psi D{\vec \Phi} \, e^{{\cal S}_{\Psi,\Phi}}\ .
\end{eqnarray}
Integrating out the fermions we obtain 
\begin{equation}
\label{Z_HSa}
\mathcal{Z}(\mu) =\tilde{\mathcal{N}} \left(\frac{\beta}{J}\right)^{3/2}\left(\frac{\beta}{J}\right)^{3N} 
\int D{\vec \Phi} \, e^{{\cal S}_{\Phi}}\ , 
\end{equation}
where the effective action ${\cal S}_{\Phi}$ depends on  $\vec \Phi_0$ and  
$\delta \vec \Phi_m$ and can be written as
\begin{eqnarray}
&& {\cal S}_\Phi =\sum_\alpha {\rm tr} \ln \beta \left[\left(G_{0,\alpha,n_1}^{-1}\right)\delta_{n_1,n_2} 
- \delta \vec \Phi_{(n_1-n_2)} \cdot {\bf S}\right] 
-\frac{\beta\Phi_0^2}{4J} - \sum_{m\neq 0} \frac{\beta\delta \vec \Phi_m \delta \vec \Phi_{-m}}{4J} \ .
\label{eq:SPhiTrLog}
\end{eqnarray}
Here $\rm tr$ stands for combined time (Matsubara frequencies) and spin trace.

\subsection{Expansion to the second order in fluctuations}
\label{appendixA2}
We expand the effective action (Eq.\ref{eq:SPhiTrLog}) up to the second order in the fluctuations 
$\delta \vec \Phi_m$. The zeroth order contribution has already been calculated in Sec.~\ref{secIII}.  
It is easy to show that the first order contribution vanishes and we obtain 
\begin{eqnarray}
&&{\cal S}_\Phi \approx const.  -\frac{\beta \Phi_0^2}{4J^\ast} 
- \sum_{m\neq 0} \frac{\beta \delta \vec \Phi_m \delta \vec \Phi_{-m}}{4J} 
-\sum_\alpha \frac{1}{2}\,{\rm tr} 
\left[G_0 (\delta \vec \Phi \cdot {\bf S})G_0 (\delta \vec \Phi \cdot {\bf S})\right]\ .
\label{eq:2ndOrderExpansion}
\end{eqnarray}
After a straightforward calculation this gives
\begin{eqnarray}
\label{eq:quadratic_term}
&&- \sum_\alpha \frac{1}{2}\,{\rm tr} \left[G_0 (\delta \vec \Phi \cdot {\bf S})G_0 (\delta \vec \Phi \cdot {\bf S})\right] =  \beta \Gamma  \sum_{m\neq 0} \frac{\delta\Phi_{-,m}  \delta\Phi_{+,-m}}{\Phi_0-i\omega_m}\nonumber\\
&&=\frac{\beta\Gamma}{2}\,\sum_{m>0}\,\frac{\Phi_0}{\Phi_0^2 + \omega_m^2}\,\left(\delta\Phi_{x.m}\delta\Phi_{x,-m}
+\delta\Phi_{y.m}\delta\Phi_{y,-m}\right)
-\frac{\beta\Gamma}{2}\,\sum_{m>0}\,\frac{\omega_m}{\Phi_0^2 + \omega_m^2}\,\left(\delta\Phi_{x.m}\delta\Phi_{y,-m}
-\delta\Phi_{y.m}\delta\Phi_{x,-m}\right)\ ,\nonumber\\
\end{eqnarray}
where $\delta \Phi_\pm = (\delta \Phi_x \pm i \delta \Phi_y)/2$ and $\Gamma$ was defined after Eq.~(\ref{Inv10}).
Similar expression was obtained earlier in Ref.~\cite{PhysRevB.77.073309} for a more involved regime of 
strong spin-orbit coupling. The last term of (\ref{eq:quadratic_term}) is purely imaginary and, at low frequencies 
(adiabatic condition $|\omega_m| \ll \Phi_0$), corresponds to the Berry phase. 
Substituting Eq.\ref{eq:quadratic_term} into Eq.\ref{eq:2ndOrderExpansion} we obtain 
\begin{equation}
\label{effS}
{\cal S}_{\Phi} =const.  -\frac{\beta \Phi_0^2}{4J^\ast}  +  {\cal S}_{\delta\Phi}\ ,
\end{equation}
where 
\begin{eqnarray}
{\cal S}_{\delta\Phi}&=& 
- \sum_{m > 0}\, \frac{\beta }{2J}\,\delta \Phi_{z,m} \delta \Phi_{-m,z} 
- \sum_{m>0}\,\left(\frac{\beta}{2J}-\frac{\beta\Gamma}{2}\,
\frac{\Phi_0}{\Phi_0^2 + \omega_m^2}\right)\, 
\left(\delta\Phi_{x.m}\delta\Phi_{x,-m} + \delta\Phi_{y.m}\delta\Phi_{y,-m}\right) \nonumber \\
&&-\sum_{m>0}\frac{\beta\Gamma}{2}\,\frac{\omega_m}{\Phi_0^2 + \omega_m^2}\,\left(\delta\Phi_{x.m}\delta\Phi_{y,-m}
-\delta\Phi_{y.m}\delta\Phi_{x,-m}\right) \ .
\label{effSdelta}
\end{eqnarray}
At this point we can perform the Gaussian integration over $\delta\vec \Phi$. We obtain
\begin{equation}
\left(\frac{\beta}{J}\right)^{3N}  \int [D{\delta \vec \Phi}] \, e^{{\cal S}_{\delta \Phi}} 
={\cal {N}^{\prime}}\prod_{m>0}^N\,\frac{\omega_m^2 + \Phi_0^2}{\omega_m^2 + (1-\nu J)^2\Phi_0^2}=
{\cal {N}^{\prime}}\,\frac{ (1-\nu J)\sinh\left[\frac{\beta\Phi_0}{2}\right]}{\sinh\left[\frac{\beta\Phi_0}{2}\,(1-\nu J)\right]} \ .
\end{equation}
where ${\cal {N}^{\prime}}$ is a $J$-independent and $\Phi_0$-independent normalization.

Thus we see that close enough to Stoner instability, i.e., for $J/J^\ast = (1 - \nu J) \rightarrow 0$, 
we reproduce Eqs.~(\ref{Inv25}) and (\ref{PFintPhia}). Indeed, this is so if  we are 
allowed to replace
\begin{equation}
\frac{ (1-\nu J)}{\sinh\left[\frac{\beta\Phi_0}{2}\,(1-\nu J)\right]} \rightarrow \frac{2}{\beta\Phi_0}
\end{equation}
for all relevant $\Phi_0$. This approximation works, at least, in the regime of relatively high temperatures, 
i.e., when $\nu^{-1} \sim J \ll T \ll J^\ast$. In this case the integral (\ref{PFintPhia}) is 
dominated by $\Phi_0 \approx J^\ast$ and we obtain $\beta\Phi_0\,(1-\nu J)\sim \beta J \ll 1$.

\subsection{Restoring the Goldstone mode in Cartesian coordinates}
\label{appendixA3}

Having already reproduced the results of the main text we would like to improve our understanding as 
well as the precision of the calculation by analyzing Eq.~\ref{effS} and Eq.~\ref{effSdelta} a bit closer. We rewrite Eq.~\ref{effSdelta}
as
\begin{eqnarray}
{\cal S}_{\delta \Phi} = 
&-& \sum_{m > 0}\, \frac{\beta }{2J}\,\delta \Phi_{z,m} \delta \Phi_{-m,z}\nonumber\\
&-&\sum_{m>0}\,\left(\frac{\beta }{2J}-\frac{\beta\Gamma}{2\Phi_0}\right)\,\left(\delta\Phi_{x,m}\delta\Phi_{x,-m}
+\delta\Phi_{y,m}\delta\Phi_{y,-m}\right)\nonumber\\
&-&\sum_{m>0}\,\frac{\beta \Gamma}{2}\,\frac{\omega_m^2}{\Phi_0(\Phi_0^2 + \omega_m^2)}\,\left(\delta\Phi_{x,m}\delta\Phi_{x,-m}
+\delta\Phi_{y,m}\delta\Phi_{y,-m}\right)\nonumber\\
&-&\sum_{m>0}\frac{\beta\Gamma}{2}\,\frac{\omega_m}{\Phi_0^2 + \omega_m^2}\,\left(\delta\Phi_{x,m}\delta\Phi_{y,-m}
-\delta\Phi_{y,m}\delta\Phi_{x,-m}\right)\ .
\label{effS1}
\end{eqnarray}
The second term of the RHS of (\ref{effS1}) is problematic since it gives a finite "mass" for the transverse fluctuations. Yet, from the spherical symmetry we should expect a Goldstone mode and no "mass".
However, the origin of this term is quite clear. It 
can be nicely rewritten as 
\begin{equation}
\label{eq:prolong}
-\frac{\beta}{4J^\ast}\sum_{m\neq 0}\,\left(\delta\Phi_{x,m}\delta\Phi_{x,-m}+\delta\Phi_{y,m}\delta\Phi_{y,-m}\right)
= -\frac{\beta}{4J^\ast} \Delta \Phi_0^2\ .
\end{equation}
Here $\Delta\Phi_0^2$ is the prolongation of the zero mode vector $\vec\Phi_0$ due to transverse fluctuations. 
This term combines with the term $-\beta \Phi_0^2/(4J^\ast)$ of (\ref{effS}) to give 
$-\beta \tilde\Phi_0^2/(4J^\ast)$, where $\tilde \Phi_0^2 \equiv \Phi_0^2 + \Delta \Phi_0^2$. Thus we "loose" the Goldstone mode 
because we use Cartesian coordinates and not the spherical ones where $\Phi_0$ would be kept constant.
This term could be just subtracted thus recovering the Goldstone physics and reducing $\tilde \Phi_0$ back to 
$\Phi_0$.  We distinguish two regimes:
\begin{enumerate}
\item[] {\textsl{(A)}} $\omega_{m}\gg\Phi_{0}$\\
At high frequencies the action (\ref{effS1}) reduces (with no subtractions) to 
\begin{eqnarray}
{\cal S}_{\delta \Phi} \approx
&-& \sum_{\omega_m \gg \Phi_0}\, \frac{\beta }{2J}\,\delta \vec \Phi_{m} \cdot \delta \vec \Phi_{-m}\ .
\end{eqnarray}
In this limit the integration over the Gaussian fluctuations produces a factor $(J/\beta)^3$ per Matsubara 
frequency, which compensates the factor $(\beta/J)^3$ per Matsubara frequency in Eq.~(\ref{Z_HS}). 
Thus it is not necessary to subtract anything at high frequencies. 

\item[] {\textsl{(B)}} $\omega_{m}\ll\Phi_{0}$\\
In this limit, after the subtraction of the second term of the RHS of (\ref{effS1}) and
retaining only transverse fluctuations we obtain
\begin{eqnarray}
{\cal S}_{\delta \Phi_\perp} = 
&-&\sum_{m>0}\frac{\beta\Gamma \omega_m}{2}\,\left(\frac{\delta\Phi_{x,m}\delta\Phi_{y,-m}
-\delta\Phi_{y,m}\delta\Phi_{x,-m}}{\Phi_0^2}\right)\nonumber\\
&-&\sum_{m>0}\,\frac{\beta\Gamma \omega_m^2}{2\Phi_0}\left(\frac{\delta\Phi_{x,m}\delta\Phi_{x,-m}
+\delta\Phi_{y,m}\delta\Phi_{y,-m}}{\Phi_0^2}\right)
\ .
\label{effSadtr}
\end{eqnarray}
In the time representation Eq.\ref{effSadtr} reads 
\begin{eqnarray}
{\cal S}_{\delta \Phi_\perp} = 
\frac{i\Gamma}{2}\,\int\limits_0^\beta d\tau\,\frac{\delta\dot \Phi_{x}\delta\Phi_{y}
}{\Phi_0^2}
-\frac{\Gamma}{4\Phi_0}\int\limits_0^\beta d\tau \,\frac{ ( \delta\dot\Phi_{x})^2 + (\delta\dot \Phi_{y})^2
}{\Phi_0^2}
\ .
\label{effSadtr_tau}
\end{eqnarray}
The first term in Eq.\ref{effSadtr_tau} is equal to the Berry phase term (\ref{Inv10}). The second term is of the second order in derivatives 
and can be rewritten as 
\begin{equation}
-\frac{\Gamma}{4\Phi_0}\int\limits_0^\beta d\tau \, (\dot {\vec n})^2\ ,
\end{equation}
where $\vec n$ is the unit vector in the direction of $\vec \Phi$. Thus we reproduce the 
second order term (\ref{Inv15}).
Since the obtained action is local in time and is proportional to the time derivatives, the fact that 
we started from an expansion in small fluctuations around a given $\vec \Phi_0$ is no longer important. 
Indeed, we can always choose the direction of $\vec \Phi_0$ close to the "current" $\vec \Phi(\tau)$. 

\end{enumerate}

\subsection{Calculation with a proper counter term}
\label{appendixA4}

As we have seen we have to subtract the "mass" term only for the adiabatic frequencies. Here we propose a particular form of the counter term which satisfies this condition. 
We rewrite again Eq.~\ref{effSdelta} as ${\cal S}_{\delta \Phi} = {\cal S}^{ren}_{\delta \Phi} + 
{\cal S}^{counter}_{\delta \Phi}$, where
\begin{eqnarray}
{\cal S}^{ren}_{\delta \Phi} 
=&-& \sum_{m > 0}\, \frac{\beta }{2J}\,\delta \Phi_{z,m} \delta \Phi_{-m,z}\nonumber\\
&-&\sum_{m>0}\,\frac{\beta}{2J}\,\frac{\omega_m^2}{\Phi_0^2 + \omega_m^2}\,\left(\delta\Phi_{x,m}\delta\Phi_{x,-m}
+\delta\Phi_{y,m}\delta\Phi_{y,-m}\right)\nonumber\\
&-&\sum_{m>0}\frac{\beta\Gamma}{2}\,\frac{\omega_m}{\Phi_0^2 + \omega_m^2}\,\left(\delta\Phi_{x,m}\delta\Phi_{y,-m}
-\delta\Phi_{y,m}\delta\Phi_{x,-m}\right)\ ,
\label{effS2}
\end{eqnarray}
and
\begin{equation}
{\cal S}^{counter}_{\delta \Phi}=-\sum_{m>0}\,\frac{\beta }{2J^\ast}\,\frac{\Phi_0^2}{\Phi_0^2 + \omega_m^2}\left(\delta\Phi_{x,m}\delta\Phi_{x,-m}+\delta\Phi_{y,m}\delta\Phi_{y,-m}\right)\ .
\end{equation}
The splitting is chosen so that ${\cal S}^{counter}_{\delta \Phi}$ represents the "mass" term at adiabatic frequencies 
but vanishes at  high frequencies. We subtract the counter term ${\cal S}^{counter}_{\delta \Phi}$. Then we obtain  
\begin{eqnarray}
\left(\frac{\beta}{J}\right)^{3N}  \int [D{\delta \vec \Phi}] \, e^{{\cal S}^{ren}_{\delta \Phi}} 
&=&{\cal {N}^{\prime}}\prod_{m>0}^N\,\frac{\omega_m^2 + \Phi_0^2}{\omega_m^2 }\,
\prod_{m>0}^N\,\frac{\omega_m^2 + \Phi_0^2}{\omega_m^2 + (J \nu \Phi_0)^2 }\nonumber\\
&=&{\cal {N}^{\prime}}\,\frac{\sinh\left[\frac{\beta\Phi_0}{2}\right]}{\frac{\beta\Phi_0}{2}} \,\left[\frac{J\nu\,\sinh\left[\frac{\beta\Phi_0}{2}\right]}{\sinh\left[\frac{\beta J \nu\Phi_0}{2}\right]}\right]\ .
\end{eqnarray}
Close to Stoner instabilty, i.e., for $J\nu \rightarrow 1$, we can approximate the second multiplier  
(in rectangular brackets) by $1$ and we reproduce the result (\ref{PFintPhia}) and (\ref{Inv29}) from 
there. In this approach no limitation on temperature from below appears.

\section{\qdd in presence of an external magnetic field \mf}
\label{appendixC}

\noindent
{\it{Here we provide an alternative derivation of the susceptibility 
by considering the isolated \qdd in a weak magnetic field. 
Unlike in the previous calculation, here the normalization factors of the path integral 
are not important.}}

In the presence of a constant magnetic field \mfd the Hamiltonian of the \qdd can be written as
\begin{eqnarray}
&&H = \sum_{\alpha,\sigma} \epsilon_\alpha  a^\dag_{\alpha,\sigma} a^{\phantom \dag}_{\alpha,\sigma} - 
J {\bf S}^2 - \vec{\mf}\cdot{\bf S}\ .
\label{mfI1}
\end{eqnarray}
Following the same procedure as described in \ref{appendixA1} we derive the effective action for this
situation. The constant magnetic field adds to the zero frequency component of the exchange field 
$\vec \Phi_0$ and, effectively, the new field $\vec{A}\equiv (\vec \Phi_{0} + \vec{\mf})$ plays the role 
of $\vec \Phi_0$ in the calculation. We obtain  
\begin{eqnarray}
&& {\cal S}_{\Phi,\mf} =\Bigg[\sum_\alpha {\rm tr} \ln \beta \left[\left(G_{A,\alpha,n_1}^{-1}\right)\delta_{n_1,n_2} 
- \delta \vec \Phi_{(n_1-n_2)} \cdot {\bf S}\right] 
-\frac{\beta\Phi_0^2}{4J} - \sum_{m\neq 0} \frac{\beta\delta \vec \Phi_m \delta \vec \Phi_{-m}}{4J} \Bigg]\ .
\label{mfI3}
\end{eqnarray}
where in $G_{A,\alpha,n}^{-1} = i\varepsilon_n -(\epsilon_\alpha - \mu) - A S_z$ is the
single particle Green's function of the electrons in presence of \mfd and 
$A=\sqrt{\Phi_{0}^{2} + \mf^{2} + 2\Phi_{0} \mf \cos\theta}$. Here $\theta$ is the angle 
between $\mf$ and $\vec \Phi_0$. 

We again expand
${\cal S}_{\Phi,\mf}$ up to the second order in fluctuations and obtain
\begin{eqnarray}
&&{\cal S}_{\Phi,\mf} \approx const.  -\frac{\beta \Phi_0^2}{4J} + \frac{\beta \nu A^2}{4}
- \sum_{m\neq 0} \frac{\beta \delta \vec \Phi_m \delta \vec \Phi_{-m}}{4J} 
-\sum_\alpha \frac{1}{2}\,{\rm tr} 
\left[G_A (\delta \vec \Phi \cdot {\bf S})G_A (\delta \vec \Phi \cdot {\bf S})\right]\ .
\label{mfI4}
\end{eqnarray}
Following the same steps as described in \ref{appendixA2} and \ref{appendixA3}
we obtain the partition function
\begin{eqnarray}
&&{\cal {Z}}=\tilde{\mathcal{N}}\left(\frac{\beta}{J}\right)^{3/2}
\int\limits_0^\infty d\Phi_0 \Phi_{0}^{2} \int\limits_0^\pi d\theta \sin\theta 
\exp{\left[-\frac{\beta \Phi_0^2}{4J}\right]}\exp{\left[\frac{\beta \nu A^2}{4}\right]}
\cdot\frac{\sinh\left[\frac{\beta A}{2}\right]}{\frac{\beta A}{2}}\
\label{mf13}
\end{eqnarray}
The susceptibility is now given by 
\begin{eqnarray}
\chi~=~-\frac{T}{{\mathcal {Z}}}\frac{\partial^2 {\cal {Z}}}{\partial \mf^2}\Big|_{\mf=0} \ .
\label{D5}
\end{eqnarray}
Performing the appropriate saddle point approximation (up to the quadratic fluctuations around 
the saddle point) we obtain the susceptibility given in Eq.~\ref{Inv29}.

%

\bibliographystyle{elsarticle-num}
\bibliography{QD_Isolated_ref}







\end{document}